\documentclass[aps,prl,twocolumn,floats,showpacs,superscriptaddress]{revtex4-1}
\usepackage{graphicx,epsfig,psfrag}
\usepackage{times}
\usepackage{graphics,dcolumn,bm,float}
\usepackage{amssymb,amsmath,rotate,color}

\usepackage{xcolor}
\usepackage{graphicx}

\begin{document}
\unitlength 1 cm
\newcommand{\be}{\begin{equation}}
\newcommand{\ee}{\end{equation}}
\newcommand{\bearr}{\begin{eqnarray}}
\newcommand{\eearr}{\end{eqnarray}}
\newcommand{\nn}{\nonumber}
\newcommand{\la}{\langle}
\newcommand{\ra}{\rangle}
\newcommand{\cd}{c^\dagger}
\newcommand{\vd}{v^\dagger}
\newcommand{\ad}{a^\dagger}
\newcommand{\bd}{b^\dagger}
\newcommand{\tk}{{\tilde{k}}}
\newcommand{\tp}{{\tilde{p}}}
\newcommand{\tq}{{\tilde{q}}}
\newcommand{\eps}{\varepsilon}
\newcommand{\vk}{{\vec k}}
\newcommand{\vp}{{\vec p}}
\newcommand{\vq}{{\vec q}}
\newcommand{\vkp}{\vec {k'}}
\newcommand{\vpp}{\vec {p'}}
\newcommand{\vqp}{\vec {q'}}
\newcommand{\bk}{{\bf k}}
\newcommand{\bp}{{\bf p}}
\newcommand{\bq}{{\bf q}}
\newcommand{\br}{{\bf r}}
\newcommand{\bR}{{\bf R}}
\newcommand{\up}{\uparrow}
\newcommand{\down}{\downarrow}
\newcommand{\fns}{\footnotesize}
\newcommand{\ns}{\normalsize}
\newcommand{\cdag}{c^{\dagger}}
\newcommand{\lc}{\langle\!\langle}
\newcommand{\rc}{\rangle\!\rangle}

\title{RKKY interaction in heavily vacant graphene}
\author{Alireza Habibi} 
\affiliation{Department of Physics, Sharif University of Technology, Tehran 11155-9161, Iran}
\author{S. A. Jafari}
\affiliation{Department of Physics, Sharif University of Technology, Tehran 11155-9161, Iran}
\date{\today}

\begin{abstract} 
Dirac electrons in clean graphene can mediate the interactions between two localized magnetic 
moments. The functional form of the RKKY interaction in pristine graphene is specified by two main
features: (i) an atomic scale oscillatory part determined by a wave vector $\vec Q$ connecting the 
two valleys. Furthermore with doping another longer range oscillation appears which arise 
from the existence of an extended Fermi surface characterized by a single momentum scale $k_F$. 
(ii)  $R^{\alpha}$ decay in large distances where the exponent $\alpha=-3$ is 
a distinct feature of undoped Dirac sea (with a linear dispersion relation) in two dimensions. 
In this work, we investigate the effect of a few percent vacancies on the above
properties. Depending on the doping level, if the chemical potential lies on the
linear part of the density of states, the exponent $\alpha$ remains close to $-3$.
Otherwise $\alpha$ reduces towards more negative values which means that the
combined effect of vacancies and the randomness in their positions makes it
harder for the carriers of the medium to mediate the magnetic interaction. 
Addition of a few percent of vacancies diminishes the atomic scale oscillations
of the RKKY interaction signaling the destruction of two-valley structure of the
parent graphene material. Surprisingly by allowing the chemical potential to 
vary, we find that the longer-range oscillations expected to arise from the
existence of a $k_F$ scale in the vacant graphene are absent. This may indicate
possible non-Fermi liquid behavior by "alloying" graphene with vacancies.
The complete absence of oscillations in heavily vacant graphene can be
considered an advantage for applications as a uniform sign of the exchange interaction
is desirable for magnetic ordering. 
\end{abstract} 
\maketitle 
\section{Introduction} 
Two-dimensional nature of graphene, along with the Dirac nature of 
charge carriers~\cite{novoselov} --as contrasted to the Schroedinger nature of the carriers in 
ordinary conductors -- makes graphene a spectacular platform for condensed matter realization
of many exciting ideas of low-dimensionality and relativistic physics~\cite{AndoReview, KatsnelsonReview}.
Among the properties of interest in materials is the nature of effective
interaction between external agents mediated by the carriers of the host 
materials. An important example of this would be the Ruderman-Kittel-Kasuya-Yosida (RKKY) 
interaction which is the exchange interaction between two magnetic impurities mediated 
via the propagators of the host. The RKKY interaction in graphene was
studied first by Saremi~\cite{Saremi} who found a generic "atomic-scale"
oscillatory behavior where the sign of the interaction alternates between the 
two sub-lattices of the bipartite honeycomb structure. The wave vector
characterizing these oscillations is $\vec Q=\vec K^+-\vec K^-$ 
where $\vec K^\pm$ denote the momenta associated with two Dirac cones.
The oscillations decay as $R^{-3}$ in long distances, where the
exponent $\alpha=-3$ comes from the linear Dirac cone dispersion in two dimensions.
The investigation of Saremi was extended by Sherafati and Satpathy in
two directions. First for the undoped graphene they extended the results
to a model of full $\pi$-bands~\cite{Sherafati1}. Secondly for the low-energy 
Dirac cone model, they considered the effect of non-zero doping~\cite{Sherafati2},
and as in ordinary two-dimensional conductors additional $k_F$-related oscillations 
with a $R^{-2}$ decay at large distances appeared. Such long-wave-length 
oscillations are due to an underlying sharp Fermi surface and disappear 
when the limit of undoped graphene ($k_F\to 0$) is approached in pristine graphene.

The above line of research was also extended to the case of disordered
graphene by Lee and coworkers who considered the effect of 
diagonal and hopping disorder on the RKKY interaction.
in undoped~\cite{Lee1} and doped graphene~\cite{Lee2}. They found that
for strong enough on-site disorders where Anderson localization takes place,
the RKKY interaction exponentially decays with distance which is 
consistent with the picture based on localized spectrum~\cite{Amini}.
For the doped graphene and weak disorder they found that the Friedel oscillations due
to the Fermi surface co-exists with the atomic-scale oscillations due
to interference between the two Dirac cones.
In this work guided by the experimentally conceivable conditions,
instead of a generic Anderson model, we consider a specific model for vacancies 
and investigate the RKKY interaction in presence of randomly distributed vacancies.
Our motivation is that various forms of defects are ubiquitous in the surface of graphene~\cite{Hashimoto}. 
The presence of defects modifies electronic, magnetic and mechanical properties of 
materials. For example vacancies generated by ion irradiation of graphene give rise to magnetic 
moments~\cite{Nair,Yazyev} and the most probable form of defects produced by irradiation 
are single vacancies~\cite{Ugeda}. Moreover coherence exchange of spin between
such magnetic moments and the spectrum of fermions of the
host graphene leads to unconventional Kondo effect in graphene~\cite{Chen_PRL,Chen_nature}.

In this work we find that the randomness caused by random vacancies
has a different physics from the Anderson model studied in Ref.~\cite{Lee1}:
We find that propagation of electron waves in defective environment reduces the
exponent $\alpha$ in $R^{\alpha}$ distance dependence of the RKKY interaction
from $\alpha=-3$ to below $-4$ away from half-filling. The exponent
$\alpha$ will not differ from $-3$ as long as the chemical potential 
is around energies where the density of states is linear in energy.
We find that $\alpha$ is controlled by the concentration of vacancies and
the chemical potential, but not much dependent on the energy level 
of the localized orbitals bound to vacancy location.
At higher concentrations of vacancies, they are 
expected to form their own impurity band, which is expected to cause a metallic 
type transport in neutral vacancies when the concentration goes beyond
a few percent~\cite{danny,allaei}. In such case we expect a Friedel
oscillations to appear as a signature of metallic state. But 
surprisingly we find that in addition to atomic-scale oscillations,
the Friedel oscillations of the Fermi surface also disappear.




\section{Model and method} 
\begin{figure} 
   \includegraphics[width=7cm]{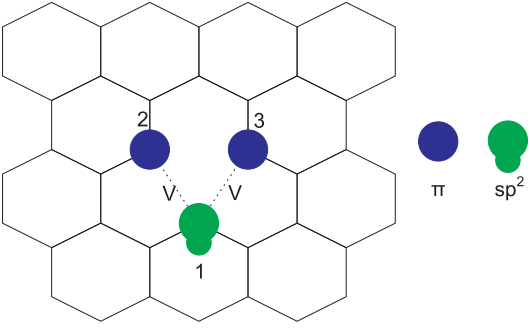} 
   \caption{(Color online) Schematic picture of the electronic state around the defect in graphene. 
   The symbol at site 1 shows the active $sp^{2}$ orbital and the circles at sites 2 and 3 the $\pi$ 
   orbitals. Two of the three $sp^{2}$ orbitals (i = 2, 3) form a covalent bond. $V$ is the amplitude 
   of the hybridization between the active $sp^{2}$ orbital and the two $\pi$ orbitals.} 
   \label{hybrid} 
\end{figure}

Kanao and coworkers proposed a simple and conceivable model for
a localized orbital and its hybridization with the adjacent 
$p_z$ orbitals needed to understand the physics of local 
magnetic moments and their interaction with the Dirac electrons
of graphene~\cite{Kanao}. They considered a Jahn-Teller distorted geometry
of a single atom point defect shown in Fig.~\ref{hybrid} where
the $sp^2$ orbital of one of the atoms, e.g. $1$ is hybridized 
with $p_z$ orbitals of the two remaining atoms $2,3$ surrounding 
the vacancy. 
The effective Hamiltonian for the defective graphene becomes~\cite{Kanao} : 
\begin{eqnarray} 
H=H_{\rm{g}}+H_{\rm{def}}+H_{\rm{hyb}} 
\label{eq_hamiltonian}, 
\end{eqnarray} 
where $ H_{\rm{g}}$ is the tight-binding Hamiltonian of graphene with 
nearest neighbor hopping and is given by
\begin{eqnarray} 
 H_{\rm{g}}=-t\sum_{\langle ij\rangle\sigma}\left(a^\dagger_{i\sigma}b_{j\sigma}+\rm{h.c.}\right)-\mu N,\label{eq_graphene_hamiltonian} 
\end{eqnarray} 
where $-t<0$ is the hopping integral and $\mu$ is the chemical potential
which can be conveniently controlled by a gate voltage. 
$H_{\rm{def}}$ is the effective Hamiltonian for $sp^{2}$ 
orbitals at sites of type $1$
\begin{eqnarray} 
 H_{\rm{def}}=\sum_{\sigma}(E_{sp^2}-\mu)d^\dagger_{\sigma}d_{\sigma},\label{eq_defect_hamiltonian} 
\end{eqnarray} 
where $E_{sp^2}<0$ is the energy level of the $sp^2$ orbital and the notation $d^\dagger_{1\sigma}$
is adopted for the creation operator at this site to emphasize its localized nature in analogy
with the situation where the localized orbital arises from a transition metal ad-atom. 
Since in this paper we are interested
in the physics of RKKY interaction, we assume that the local moments are formed,
and concentrate on their exchange interaction mediated by the electronic states of
the defective host. Therefore we do not include the on-site Coulomb repulsion 
between the electrons in this localized orbital. 
Finally $H_{\rm{hyb}}$ is the hybridization which is given for each defect by: 
\begin{eqnarray} 
   H_{\rm{hyb}}=V\sum_{\sigma=\uparrow\downarrow}
   \left[\left(a^\dagger_{2\sigma}+a^\dagger_{3\sigma}\right)d_{1\sigma}+\mathrm{h. c. }\right]
   \label{eq_hybridization}, 
\end{eqnarray} 
Here $a_{i\sigma}$ for $i=2,3$ is an annihilation operator of a $p_z$ electron at site $i$ with 
spin $\sigma=\uparrow,\downarrow$ and $V$ is the amplitude of hybridization. 
The missing carbon atom can belong to both sub-lattices. The Jahn-Teller distortion
preferring atom $1$ over the other two can take place in any direction. In this paper
we take all three possibilities with equal probability.
\begin{figure} 
   \includegraphics[width=8.5 cm]{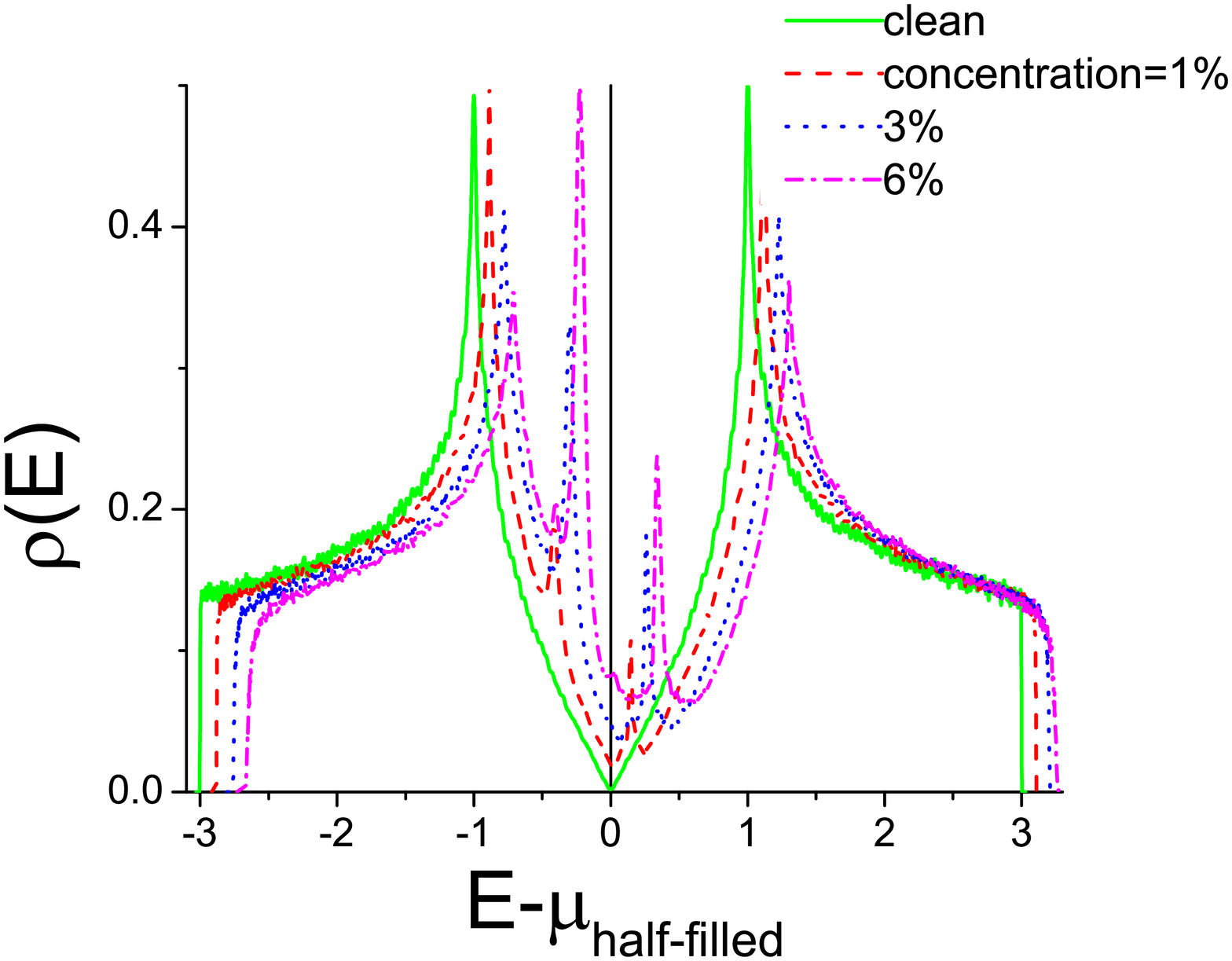} 
   \caption{(Color online) Plot of the density of states $\rho(E)$ as a function of $E-\mu$. 
   By increasing vacancy concentration, $\mu_{\textrm{half-filled}}$ 
   be shifted to the left of the Dirac neutrality point as the acceptor band
   to the left of the Dirac point becomes stronger and accommodates more electrons.} 
   \label{dos-mu} 
\end{figure} 
In Eq.~\ref{eq_hybridization}, 
we set $V=-0.2t$ for the amplitude of hybridization between $p_z$ and $sp^2$ 
orbitals~\cite{Porezag} and $E_{sp^{2}}=-0.5t$, unless otherwise specified.

To have a feeling for what happens to the density of states (DOS) of 
the vacancy alloyed graphene, in Fig.~\ref{dos-mu} the DOS of clean honeycomb 
lattice has been compared to highly vacant graphene. By increasing the
concentration of defects, two peaks to the left and right
of the Dirac point develop which can be considered as the
analogue of bonding and anti-bonding states formed due to the
Hybridization term. The chemical potential can be tuned with 
a gate voltage. In the absence of a gate voltage and for
neutral vacancies we have the half-filled situation corresponding
to one $2p_z$ electron per existing carbon atom. The chemical 
potential corresponding to half-filling deviates to the left of
zero when the defects are introduced to the graphene sample. 
Therefore the $\mu=0$ in the vacant graphene corresponds to 
electron doping. 

The general perturbative expression for the RKKY exchange interaction 
is given by~\cite{Solyom,white},
\be
   J_{\textrm{RKKY}}  =  \frac{J^2S(S+1)}{4 \pi S^{2}} \int d\omega  
   f(\omega) \textrm{Im} \left[ G( {\vec r_j} ,{\vec r_i}, \omega) 
   G(\vec{r}_i,\vec{r}_j,\omega)\right] \nn
\ee
where $S$ is the magnitude of the impurity spin, $J$ is the interaction between the localized moments and the spin of the itinerant electrons, $i$ and $j$ are the site index of magnetic impurities which are located at position $\vec{r}_{i}$ and $\vec{r}_{j}$, $ f(\omega)$ is the Fermi-Dirac distribution function. 
Ignoring the constants appearing before the integral, the integral appearing in this 
equation for our purposes can be simplified to,
\be
\Im \int\!\! d\omega f(\omega) 
\sum_{n,m} \frac{F^{ij}_{nm}}{(E_n - \omega +  i\delta)(E_m - \omega + i\delta)}. 
\label{eq:RKKY_original} 
\ee
where $F^{ij}_{nm} = \psi^*_n(\vec{r}_i) \psi_n(\vec{r}_j) \psi^*_m(\vec{r}_j) \psi_m(\vec{r}_i)$, 
with $\psi_n(\vec{r}_i)$ denoting the eigenvector corresponding to the eigenvalue $E_n$ of the 
host Hamiltonian. The lattice constant $a$ and $\hbar$ are set to unity in all calculations. 
We assume that temperature is zero ($T=0$) so that the Lehman representation of the above 
integral in terms of appropriate spectral functions becomes,
\begin{equation} 
    J_{\textrm{RKKY}} = -\int_{\eps<\mu} d\eps \int_{\eps'>\mu} d\eps' 
    \frac{F(\eps,\eps')}{\eps-\eps'}, 
    \label{eq:J-KPM} 
\end{equation} 
where $F(\eps,\eps') = \textrm{Re}[A_{ji}(\eps)A_{ij}(\eps')]$ and the real-space "spectral function" 
is given by $A_{ij}(\eps) = \langle i|\delta(\eps-H)|j\rangle$. 
The kernel polynomial method (KPM) can be conveniently used to evaluate various
spectral functions including $A_{ij}(\eps)$ ~\cite{Lee1,Amini,Fehske}. 
In this method, $A_{ij}(\eps)$ can be expressed as a series in Chebyshev 
(or any other complete set of orthogonal) polynomials and the expansion coefficients 
-- the so called moments -- are evaluated by repeated operation with powers of the 
appropriate re-scaled Hamiltonian in such a way that a $T_n(H)$ is generated 
via their recursive relations. In the appendix B we briefly summarize the KPM method.
The defects considered here are on the scale of few percent and their positions in the
lattice as well as the preferred direction due to Jahn-Teller distortion  
is drawn from a uniform random distribution. 
Typical averages are obtained by geometric averages of the form
$J_{\rm RKKY}^{\rm geo}=\exp(\langle (1/2) \ln (J_{\rm RKKY})^{2}\rangle_{\rm ave})$.
The geometric averages of the above type are known to better represent the propagating
nature of waves, and when they become zero indicate the transition to insulating 
state~\cite{Lee1,Lee2}. To ensure the same holds for our particular form of disorder, 
in appendix A we have compared the statistics of geometric and arithmetic averages.

\section{Numerical results for the RKKY exchange}
\subsection{RKKY interaction in clean graphene} 
\begin{figure} [b]
   \includegraphics[width=8 cm]{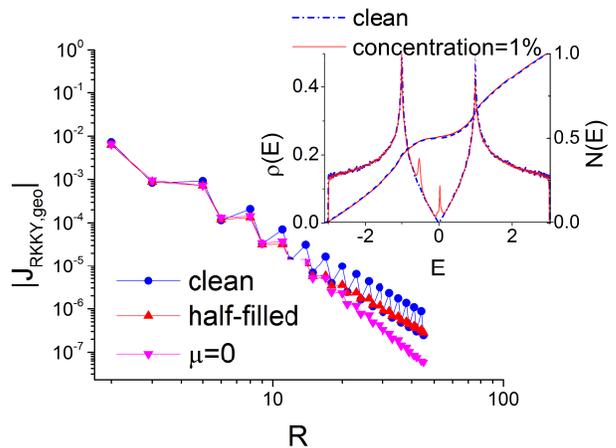} 
   \caption{(Color online) Plot of the $J_{\textrm{RKKY}}$ as a function as $R$ in the armchair
   direction. By increasing distance $R$ between two magnetic impurities an atomic-scale sign
   alternation of $J_{\textrm{RKKY}}$ agrees with the analytic formula~\cite{Sherafati1}. 
   For a $1\%$ concentration of vacancies the oscillations due to interference between the
   two valleys are washed out. The change in the exponent $\alpha$ depends on the chemical
   potential.}
   \label{J_R_armchair_C} 
\end{figure}

Let us first discuss the performance of KPM in addressing the RKKY interaction.
In Ref.~\onlinecite{Sherafati1}, the authors used a lattice Green's function method 
to calculate the RKKY interaction for pristine graphene beyond the low-energy model
of Dirac fermions. Their result agrees with other authors~\cite{Saremi,Black-Schaffer}. 
They obtain the following analytical results for the clean honeycomb lattice: 
\begin{eqnarray} 
   J_{\textrm{AB}}^{0} & = & 3C J^2\frac{1+\cos[\vec Q\cdot {\vec R}+\pi-2\theta_{R}]}{(R/a)^3} 
   \label{eq:rkky_dirac_AB} \\
   J_{\textrm{AA}}^{0} & = & - C J^2\frac{1+\cos[\vec Q\cdot\vec{R}]}{(R/a)^3} 
   \label{eq:rkky_dirac_AA} 
\end{eqnarray} 
where $C=9\lambda^2 \hbar^2/(256\pi t)$ is a positive quantity, $a$ is the lattice 
constant, ${\vec K}^{\pm}=(\pm 2\pi/(3\sqrt{3} a),2\pi/(3 a))$ are the Dirac points in the 
momentum space, ${\vec {R}}={\vec{ r}}_i - {\vec{ r}}_j$, and angle $\theta_R$ is the polar 
angles between $\vec{R}$ and $\vec Q=\vec{K^{+}}-\vec{K^{-}}$. 
These equations indicating  ferromagnetic (antiferromagnetic) correlation on same (different) sub-lattice
were numerically confirmed with the KPM method~\cite{Lee1,Lee2}. We have also checked that the KPM
results obtained in the clean limit (see Fig.~\ref{J_R_armchair_C}) agree with the 
numerically exact method of Green's functions for the pristine graphene~\cite{Sherafati1,Lee1}. 
The strength of the KPM method is that it can handle
clean and disordered systems on the same footing. For the disordered case only an averaging over
realizations of disorder will be required which is what we explain in detail in the following subsection.

\subsection{RKKY interactions in heavily vacant graphene} 
We create evenly distributed defects on the graphene in order to 
measure their effect on the mediation of RKKY interactions. 
The energies are measures in units of the
hopping amplitude, i.e. we have set $t=1$. We used a graphene sheet with periodic boundary condition 
and $2.5 \times 10^5$ lattice points are used. We generate $1.5\times10^3$ different 
random configurations for the distribution of vacancies. 
The number of Chebyshev moments is in the range $3000-4500$. 
The hybridization parameter is fixed at $V=-0.2t$, and $E_{sp^2}=-0.5t$.
 
In Fig.~\ref{J_R_armchair_C} we have plotted the distance dependence of the RKKY interaction
for a sample with $1\%$ vacancies. The inset shows DOS and the integrated DOS (electron number)
for clean and $1\%$ vacant sample. The results are reported for the armchair direction 
The first important feature in pristine graphene is the presence of atomic
scale oscillations which are due to the interference between two valleys.
The second feature which has become more manifest in the log-log plot is 
that the atomic scale oscillations diminish in long distances
for both $\mu_{\rm half-filled}$ and $\mu=0$ cases. 
This means that the Brillouin zone scale wave vector connecting the two Dirac cones of the
pristine graphene does not exist in $1\%$ vacant sample anymore. 

Another feature seen in the
two plots corresponding to vacant sample is that the slope of the exponent $\alpha$
characterizing the overall power-law decay of RKKY interaction
stays at $\alpha=-3$ for the half-filled situation, while it is below $\alpha=-3$ for the
electron doped ($\mu=0$) case. To further explore this observation, we check that when  
the chemical potential falls in the $\rho(E)\propto|E|$ region (which is naturally inherited 
from the density of states of graphene parent) the exponent remains at
$\alpha=-3$ in agreement with other KPM study~\cite{Lee2}. 
Indeed a linear average DOS in two dimensions corresponds to a linear dispersion 
relation, which by dimensional analysis is expected to give rise to $R^{-3}$ dependence in the 
particle-hole bubble. 
This explains why in low concentration regime, despite introducing 
$1\%$ vacancies, the exponent $\alpha=-3$ in Fig.~\ref{J_R_armchair_C}. 
With this point in mind, we now proceed to study
the variations in the exponent $\alpha$ in vacant graphene at $\mu=0$.

\subsection{Doping dependence} 
\begin{figure} 
 \includegraphics[width=8.5 cm]{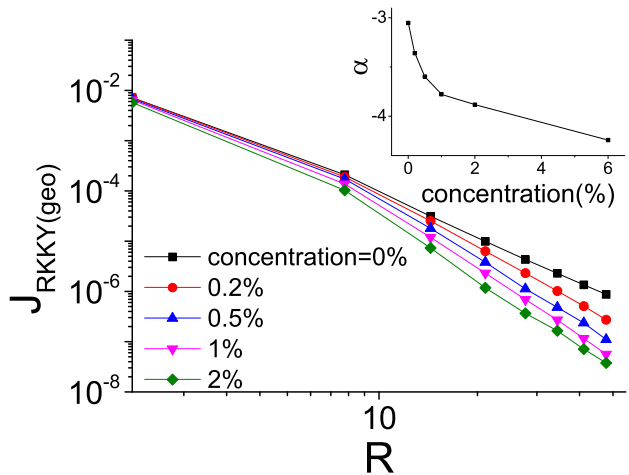} 
\caption{(Color online) (a) Plot of the $J_{\textrm{RKKY}}$ as a function as $R$ in the
electron doped situation ($\mu=0$). By increasing distance between two magnetic impurities in 
the armchair direction, $J_{\textrm{RKKY}}$ still decreases with by a power-law. 
(b) Plot of the $\alpha$ as a function of vacancy concentration in armchair direction. 
By increasing vacancy concentration, $\alpha$ will decrease. 
For the zigzag direction the trend is similar, with slightly different 
exponents at a given concentration.
} 
\label{J_R_armchair_VC_alpha_concentration_armchair} 
\end{figure}

Based on the discussion in previous subsection, at $\mu=0$ which corresponds to 
electron doping in vacant graphene and the chemical potential falls in 
the region of DOS which is strongly nonlinear, we expect significant variations in the 
exponent $\alpha$ which can be attributed to hindered motion of electron waves inside
the vacant graphene medium. For this, in Fig.~\ref{J_R_armchair_VC_alpha_concentration_armchair}
we plot the dependence of RKKY interaction to distance in a log-log scale for the 
armchair direction. To extract the exponent $\alpha$ in the clean and vacant systems,
first we discard the length scales below $R=10a$, as the power-law is meant for long
distances. Secondly, slight oscillations surviving the presence of disorder are removed
in order to remain focused on the power-law part of the dependence. We will return to
the question of oscillations in the sequel. 

The clear change in the exponent $\alpha$ indicates that the propagation of 
electrons in heavily vacant graphene is harder than the clean sample which 
is conceivable as the disorder is expected to make the propagation of
both single-particle and particle-hole excitations more difficult. 
Based on dimensional arguments, the change in $\alpha$ could be effectively
ascribed to a change in the average dispersion relation of the electronic
states characterizing the vacant sample. 
Along the zigzag direction a similar conclusion holds, with a minor difference
that for a given concentration, the value of the exponent $\alpha$ for
the zigzag direction slightly differs from the armchair direction. 
The qualitative distinction between the armchair and zigzag directions of the
pristine graphene does not prevail to the realm of high concentration of 
vacancies and the qualitative behavior along the two directions are not expected
to be much different when a few percent vacancies are introduced. 

\begin{figure}[h] 
\includegraphics[width=8.5 cm]{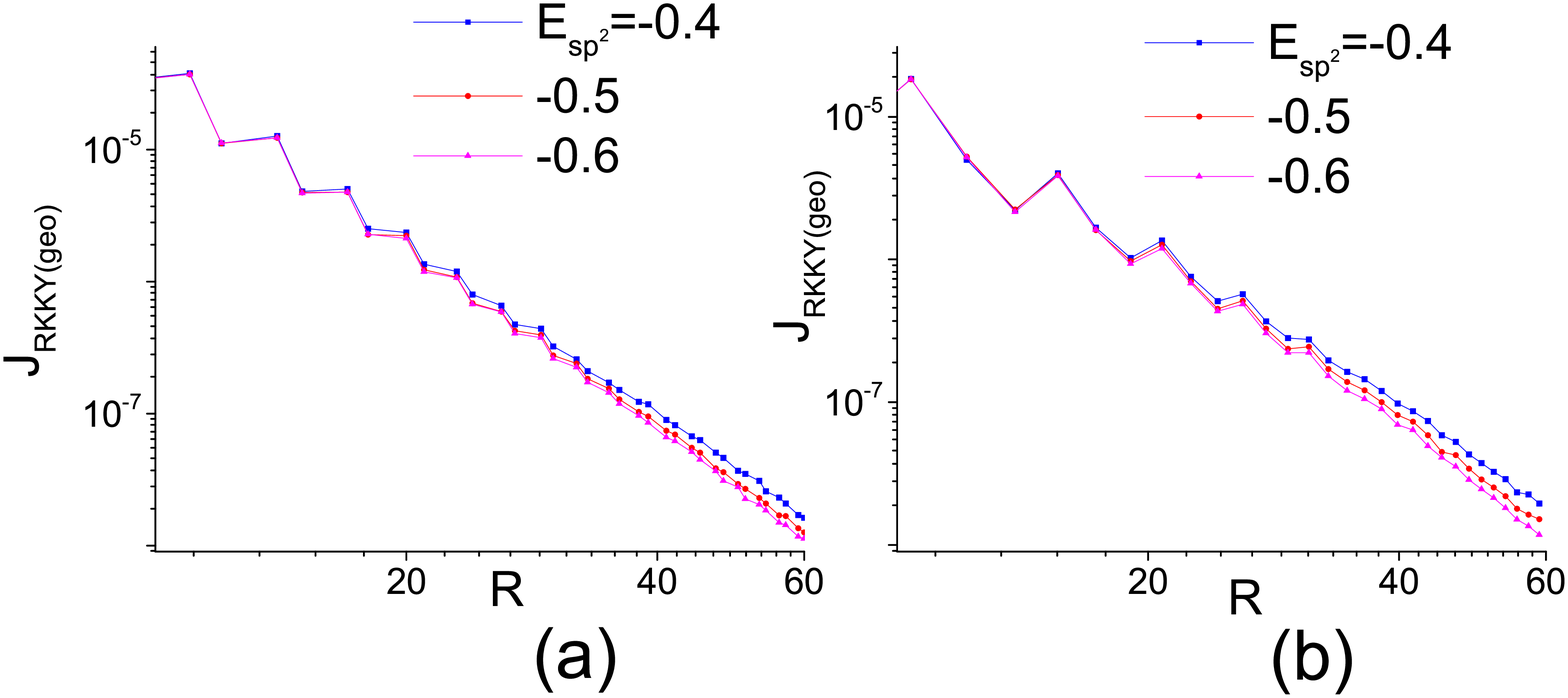}
\caption{(Color online) Plot of the $J_{\rm RKKY}$ as a function as distance for (a) armchair and (b) 
zigzag directions at $\mu=0$ for different values of $E_{sp^2}$.  
As can be seen, the exponent $\alpha$ is not much 
sensitive to the precise value of $E_{sp^2}$.} 
\label{J_R_armchair_and_zigzag_Esp2} 
\end{figure}

At the $\mu=0$ situation we also examine the dependence of $\alpha$ on the 
energy level $E_{sp^2}$ of the vacancy. As can be seen in Fig.~\ref{J_R_armchair_and_zigzag_Esp2}
this exponent is not sensitive to the precise value of the impurity level as long
as it is not deep enough to allow for the formation of localized bound states which
totally changes the nature of propagation of electronic waves. 

\subsection{The fate of Fermi-surface oscillations}
\begin{figure}[b] 
\includegraphics[width=8.0 cm]{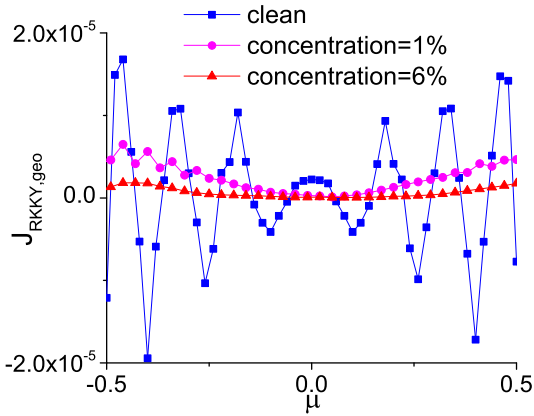} 

\caption{(Color online) Plot of the $J_{\textrm{RKKY}}$ as a function as $\mu$ for 
armchair direction ($R=32a$). Vacancies tend to wash out Friedel oscillations.
More vacancies do a better job.
 } 
\label{mu_J_R3_armchair} 
\end{figure} 
When the pristine graphene is doped, in addition to atomic scale oscillations in the
RKKY interaction which merely arise from the wave vector $\vec Q$ connecting the two
Dirac cone valleys, there appears another longer range oscillations due to the
presence of extended Fermi surface with characteristic wave-vector $k_F$~\cite{Sherafati2}.
When the disorder is introduced within the Anderson model, both atomic scale
oscillations and the Friedel oscillations due to Fermi surface survive~\cite{Lee2}. 
Note that the Dirac picture remains quire robust against the weak disorder~\cite{Amini}.
Therefore given the dimensional consistency the overall physics of the RKKY interaction
is expected to survive in weak disorders considered in~\cite{Lee2}.
As we show in the following, for the model of vacant graphene considered here,
even the Fermi surface oscillations are washed out. To begin with, note that
since the functional dependence of the oscillatory part of the RKKY interaction on 
distance $R$ always comes through a combination $k_FR$, it could be viewed 
as an oscillatory function of $k_F$ instead. With this in mind, in 
Fig.~\ref{mu_J_R3_armchair} we have plotted the $\mu$-dependence of the pristine 
and vacant graphene with $1\%$ and $6\%$ vacancies at a fixed distance $R=32a$.
For the clean graphene first of all a perfect particle-hole symmetry is seen in
the RKKY interaction which is a symmetry of the Hamiltonian. Secondly an oscillatory 
dependence on $\mu$ is seen which implies oscillatory dependence on $k_F$ and hence
on $k_FR$ on dimensional grounds. 
By adding vacancies to the system note that the particle-hole symmetry is 
lost as the presence of vacancies breaks such symmetry. Now at $0.01$ vacancy
concentration the there are no Friedel oscillations up to $\mu$ values where
the pristine graphene would complete one cycle of oscillations. Some oscillations
with reduced amplitude can be seen below the $\mu\approx-0.3$. 
When the vacancy concentration is increased up to $0.06$ no Friedel oscillations 
can be found. 
\begin{figure}[t] 
\includegraphics[width=8.5 cm]{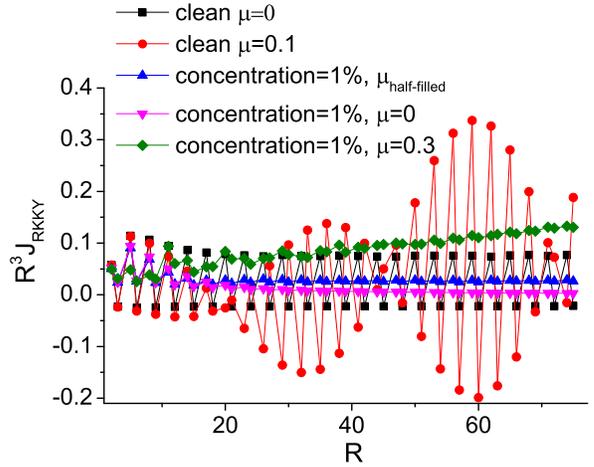} 
\caption{(Color online) Plot of the $R^3J_{\textrm{RKKY}}$ as a function as $R$ for armchair direction. 
The clean limit is compared with $1\%$ vacancy situation at different doping levels.
Adding one percent vacancies kills both type of short-ranged and long-ranged oscillations
for the considered chemical potentials.
} 
\label{R_J_R3_armchair} 
\end{figure} 

To compare the real-space profile of the oscillations more clearly, 
in Fig.~\ref{R_J_R3_armchair} we have plotted the $R^3J_{\rm RKKY}$
as a function of $R$ along the armchair direction for various values of
$\mu$ at vacancy concentration $0.01$. The plots are contrasted to the 
case of pristine graphene. As can be seen in the pristine graphene, 
when the chemical potential is at $\mu=0$ which corresponds to undoped
case, the function $R^3J_{\rm RKKY}$ only shows an atomic scale oscillations at
large $R$ (black, filled-square plot). When the chemical potential is tuned
away from the Dirac point in pristine graphene (the red, filled-circle plot)
it can be clearly seen that in addition to the atomic scale oscillations, 
a Friedel type oscillations due to the presence of a Fermi surface with a
definite characteristic scale $k_F$ appear and are superimposed into the
atomic scale oscillations in agreement with analytical results for clean graphene~\cite{Sherafati2}.
However the striking observation is made when one percent vacancies are introduced. 
In such case, not only the atomic scale oscillations are diminished, but also
no sign of oscillations due to "Fermi surface" are seen at chemical potentials
corresponding to half-filling and $\mu=0.1$. At $\mu=0.3$ it seems that some tiny
atomic scale oscillations are present, while longer range Fermi-surface oscillations
are still absent. The absence of Friedel-type oscillations in vacant graphene 
indicates that in presence of such amount of vacancies the picture of a Fermi 
surface with a sharp length scale $k_F^{-1}$ does not hold anymore. It is reminiscent
of disorder driven non-Fermi-liquid behavior in Kondo alloys~\cite{Miranda} where 
the creation of vacancies plays a role akin to alloying. Therefore it is likely that
in such case the Fermi surface is characterized with multitude of length scales
arising from local Fermi surfaces, the disorder averaging over which washes out the 
part of oscillations corresponding
to presence of a sharp single Fermi wave-vector scale of clean metallic state. 
It should also be noted that in Fig.~\ref{R_J_R3_armchair} in the case of pristine
graphene the amplitude of long-range oscillations in the function $R^3J_{\rm RKKY}$
increases which indicates the decay power in these type of terms is weaker than the
$R^{-3}$ behavior~\cite{Sherafati2}. The same argument for the vacant system at
$\mu=0.3$ indicates that the decay power of the $J_{\rm RKKY}$ is not as strong as
$R^{-3}$ behavior. 

To summarize we find that when vacancies on the scale of a percent are
introduced to graphene, the nature of RKKY interaction changes as follows:
when the doping level is such that the chemical potential falls in the energy
region where DOS is linear, the exponent $\alpha$ in $R^{\alpha}$ remains
at $\alpha=-3$. Otherwise presence of vacancies pushes $\alpha$ to more
negative values. Regarding the oscillatory nature of the RKKY interaction
in pristine graphene, a few percent vacancies wash out both atomic scale 
oscillations and the Friedel oscillations due to Fermi surface for a remarkable
range of chemical potential values. 

\section{Acknowledgement}
SAJ thanks the school of physics, Institute for fundamental research (IPM)
for support.

\section{Appendix}
\appendix

\section{A: Statistics of geometric averages} 
\begin{figure}[t] 
\includegraphics[width=8.5 cm]{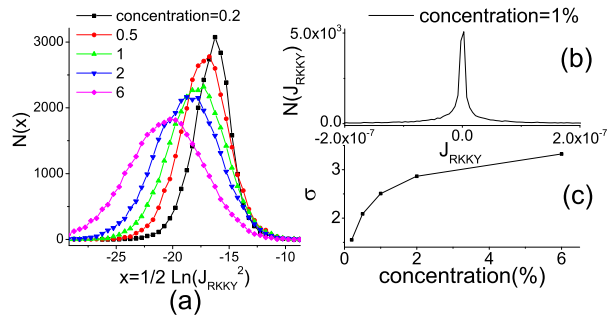}
\caption{(Color online). Details of the statistics. (a) The distribution of the quantity 
$x=(1/2) \ln{(J_{\textrm{RKKY}}^2)}$ sampled from the armchair direction distance $R=60 a$. 
(b) The non-Gaussian distribution of $J_{\textrm{RKKY}}$ (corresponding to arithmetic averaging). 
(c) Plot of $\sigma$ in geometric averaging as a function of the percentage of vacancies 
in armchair direction. By increasing the concentration of vacancies the variance will increase. 
All data are evaluated for $3\times10^4$ configurations.
} 
\label{J_R_arm_sigzag_histo} 
\end{figure}

In this appendix we present the statistics of our geometric averaging process.
We evaluate the probability distributions for both
$x=(1/2) \ln{(J_{\textrm{RKKY}}^2)}$ (corresponding to geometrical averaging) 
and $J_{\textrm{RKKY}}$ (corresponding to arithmetic averaging) in 
Fig.~\ref{J_R_arm_sigzag_histo} panels (a) and (b) respectively. 
As can be seen the distribution in $x$ is a Gaussian,
\begin{equation} 
 P(x)=\dfrac{N}{\sqrt{2 \pi \sigma^2}} \exp{\left[-\dfrac{(x-x_0)^2}{2\sigma^2}\right]}. 
\label{eq:J-prob} 
\end{equation} 
By fitting the Gaussian to the distribution data in panel (a), we find the corresponding
width $\sigma$ of the distribution which as been plotted in Fig.~\ref{J_R_arm_sigzag_histo} (c) 
as a function of the impurity concentration. It is interesting to compare the
behavior in Fig.~\ref{J_R_arm_sigzag_histo}b with the linear behavior reported in 
Ref.~\cite{Lee2} for a the Anderson model disorder. 

\section{B: Adjusting the chemical potential in KPM} 
In this appendix we derive a simple relation that enables us to 
find out the chemical potential corresponding to a given number
of electrons in the system. To be self contained, we briefly review
the kernel polynomial method (KPM) as well~\cite{Fehske}.

Consider a quadratic Hamiltonian $H$ whose energy eigenvalues $E$ are
limited to a bandwidth $[E_{min},E_{max}]$. Rescaling the Hamiltonian from 
$H(E)$ to  $\hat{H}(\varepsilon)$ where $\hat{H}=(H-b)/a$ and 
$\varepsilon=(E-b)/a$ where $b=(E_{max}+E_{min})/2 $ and  $a=(E_{max}-E_{min})/2$;
one can expand spectral functions such as the density of states (DOS) in a complete
set of e.g. Chebyshev polynomials defined in the range $\eps\in[-1,1]$ as,
\begin{eqnarray} 
\hat{\rho}(\varepsilon)=\frac{1}{\pi \sqrt{1-\varepsilon^2}}(\mu_0 g_0+2\sum_{m=1}^{N_c}\mu_m g_m T_m(\varepsilon) )
\label{dos} 
\end{eqnarray} 
where $T_m(\varepsilon)=\cos(m \arccos(\varepsilon))$  are the $m$th Chebyshev polynomials, 
$\mu_m$ are Chebyshev moments and $g_m$ are appropriate attenuation factors to minimize
the Gibbs oscillations. The moments are traces of polynomials of the Hamiltonian and can be 
statistically calculated as a trace, 
$\mu_m=1/r \sum_{r=1}^{M} \langle\phi_r\vert T_m(\hat{H}) \vert \phi_r \rangle$, where 
$\phi_r$ are a random single particle states and $M$ is the number of random realizations 
used in numerical calculations and $N_c$ is a large number where the expansion is cut off. 
To obtain the effect of $ T_m(\hat{H})$ on a given ket, the recurrence relation of Chebyshev 
polynomials, $T_m (\hat{H})= 2\hat{H} T_{m-1}(\hat{H}) - T_{m-2}(\hat{H})$ with 
initial conditions $T_{1}(\hat{H})=\hat{H}$ and $T_{0}(\hat{H})=1$ are used. 
At the end the DOS in the original bandwidth scale is obtained as $\rho(E)= \hat{\rho}(\varepsilon)/a$. 

The spectral functions corresponding to space non-diagonal propagation can also be similarly 
calculated between any two points $i$ and $j$ in the lattice,
\begin{eqnarray} 
\hat{\rho}_{ij}(\varepsilon)=\frac{1}{\pi \sqrt{1-\varepsilon^2}} (\mu_0^{ij} g_0+2\sum_{m=1}^M \mu_m^{ij} g_m T_m(\varepsilon) )
 \label{rho_ij} 
\end{eqnarray} 
where the non-local moment $\mu_m^{ij}$ is given by the matrix element 
$\langle i \vert T_m(\hat{H}) \vert j \rangle$. 

When we add vacancies in a graphene sheet, the particle-hole symmetry is lost,
and the half-filling will not correspond to $\mu=0$ anymore. Therefore within the
KPM we need to find a formula to give a relation between the particle number density $n$ and
the chemical potential $\mu$. The filling factor is given by,
\begin{eqnarray} 
n(\mu)=\int_{-1}^{+1}  \theta(\varepsilon-\mu) \hat{\rho}(\varepsilon) d\varepsilon 
 \label{electron_number1} 
\end{eqnarray} 
where $\theta(\varepsilon-\mu)$ is Heaviside function and $\mu$ is the (scaled) Fermi level.
Expanding the integral in Chebyshev polynomials and using the trigonometric representation
of Chebyshev polynomials one can easily perform the integration to get, 
\be
   n\!=\!\frac{\mu_0 g_0}{\pi} \arccos(-\mu)
   -2\sum_{m=1}^{N_c} \frac{\mu_m g_m} {m\pi} \sin(m\arccos\mu)
   \label{electron_number_cheb} 
\ee
For a given $n$ one needs to adjust $\mu$ to satisfy this equation. Then the
physical $\mu$ is obtained by undoing the rescaling. For example 
it can be checked that in clean graphene, the solution of the above equation 
with $n(\mu)=0.5$ will be $\mu=0$.

 
\end{document}